# Electrical control of superposed quantum states evolution in quantum dot molecule by pulsed field

S. W. Hwang*†, D. Y. Jeong*, M. S. Jun*†, M. H. Son*, L. W. Engel‡, J. E. Oh§, & D. Ahn*

**Institute of Quantum Information Processing and Systems, University of Seoul, 90 Jeonnong, Tongdaemoon-ku, Seoul 130-743, Korea*

† *Department of Electronics Engineering, Korea University, Anam, Sungbuk-ku, Seoul 136-075, Korea*

‡*National High Magnetic Field Laboratory, University of Florida, Tallahassee, Florida 32310, U. S. A*

§*Department of Electronic Engineering, Hanyang University, Ansan, Kyunggi-do 425-791, Korea*

**Quantum computing (or generally, quantum information processing) is of prime interest for it potentially has a significant impact on present electronics and computations[1-5]. Essence of quantum computing is a direct usage of the superposition and entanglement of quantum mechanical states that are pronounced in nanoscopic particles such as atoms and molecules[6, 7]. On the other hand, many proposals regarding the realization of semiconductor quantum dot (QD) qubits[8,9] and QD quantum gates[10] were reported and several experimental clues were successfully found[11-16]. While these pioneering experiments opened up the basis of semiconductor QD quantum computing, direct electrical control of superposed states in time domain[17] has not been realized yet. Here, we demonstrate the coherent control of the superposed quantum states**

**evolution in the QD molecule. Short electrical pulses are shown to inject single electron into the QD molecule and the current collected at the substrate exhibits oscillations as a function of the pulse width. The oscillations are originated from different decay rates of the symmetric (S) and the anti-symmetric (AS) state of the QD molecule and the evolution of the occupation probabilities controlled by the injection pulse width.**

The QD molecule was realized by a molecular beam epitaxy (MBE) growth of double layers of self-assembled InAs QDs stacked with the separation of 5 nm (transmission electron micrograph (TEM) image of Fig. 1a). The controllability of the dot size, the uniformity, and the stack separation was proven in the previous works[18]. The S/AS splitting of our QD molecule is evident by the magnetic field (B) spectroscopy obtained from the transport measurement of a n-i-n tunneling structure (Fig. 1b)[19]. The n-i-n structure incorporates two stacked layers of InAs QDs embedded in the GaAs well between AlAs barriers. The differential conductance (dI/dV) exhibits double peaks and they are split into four peaks at high B. This is contrasting to the usual single dot transport that shows single peak splitting into two peaks at high B[20]. The splitting between the S and the AS states ($\Delta$Es) estimated from the voltage spacing between two peaks at B = 0 is ~ 1 meV. The upper bound of the charging energy estimated from the position of S state peak is ~ 9 meV. The same stacked dot structure inside an undoped GaAs layer was used for the pulse measurement device with Au pulse electrodes on top (Fig. 1c). The $n^+$-type substrate collects the electrons injected by the pulse electrode that are impedance-matched with a superconducting coaxial cable. The DC current-voltage (I-V) and dI/dV-V of the pulse device show double peaks that are similar to the case of the n-i-n structure (Fig. 1d). The peak in dI/dV occurs when the chemical potential of the electrode is aligned with the S and AS state of coupled QDs. Assuming parabolic band-bending in the GaAs region[21, 22], $\Delta$E from the position of two peaks is estimated to be also ~ 1 meV. This value is consistent with one obtained from the theory[23].

The pulsed I-V was measured by applying a series of voltage pulses with the duration of Δt and the repetition rate of 100 MHz at the pulse electrode. During a long-term repetition (order of seconds) of the applied pulses, the average (DC) substrate current ($I_{sub}$) was monitored. The effective magnitude of the voltage pulse ($V_{eff}$) at the sample was measured by the sampling scope with the same wave-guide structures. Figure 1e shows the $I_{sub}$-$V_{eff}$ when Δt = 100 psec. Similar structures as in the case of DC data are reproduced while the magnitude of $I_{sub}$ is much smaller than the DC current.

The controlled superposition of S/AS states by the pulsed injection of electrons is demonstrated by the detailed measurement of $I_{sub}$ as a function of Δt at a fixed $V_{eff}$. The injected electrons into the QD molecule will occupy either S or AS state and the relative occupation probability changes periodically as a function of Δt (schematic in Fig. 2a). At the end of the pulse, the system goes back to the thermal equilibrium and the coherent evolution between two states stops. Then the injected electron decays into the substrate and is collected by the current preamplifier. The decay rate is larger for the AS state and $I_{sub}$ will be larger when the occupation probability of AS is larger. Figure 2b shows $I_{sub}$ vs Δt and $dI_{sub}/d(\Delta t)$ vs Δt when $V_{eff}$ = - 0.7 V (corresponding to the middle of S and AS state peaks). First of all, a wide plateau (50 psec < Δt < 350 psec) followed by a strong increase of $I_{sub}$ is observed. Secondly, three small staircases in the large plateau are observed in $I_{sub}$ as well as in $dI_{sub}/d(\Delta t)$. Figure 3a – 3c show the oscillation of $I_{sub}$ after removing the background ($\Delta I_{sub}$) as a function of Δt taken at the positions denoted by A - C in Fig. 2b. The oscillations as a function of Δt are clearly notified and the periodicity changes from 4 to 5 psec as shown by their corresponding Fourier transform (the resolution of Δt is 1 psec). These periodicities correspond to ΔE of 1 meV in the QD molecule ($h/\Delta E$) that is identified in the DC transport in Fig. 1. Figure 3d shows $\Delta I_{sub}$ of the region C at the temperature (T) of 88 K ($\Delta E << k_B T$), which does not show any clear oscillations. Even at this high T, the large plateau and the staircases persist. Figure 3e shows $\Delta I_{sub}$ exhibiting clear transition between the region C and the region where $\Delta t > 400$ psec (The Fourier transform in the right side of Fig. 3e is for the window $400 < \Delta t < 450$ psec and it does not show any clear peak). Finally, we did not observe any clear oscillations at the second staircase (in between A and B in Fig. 2b).

The current through the QD molecule is limited by the decay (discharge) time from the QD since the injection (charging) time is much shorter than the decay time. The DC current per single QD molecule at V = - 0.7 V estimated from the measured I of 0.2 μA, the active area of injection ($A_{dot} = 50\times50\ \mu m^2$), and the areal density of the QD molecules ($N_{dot} = 5\times10^{10}\ cm^{-2}$) is $I/(N_{dot}A_{dot}) = 0.16$ pA/molecule. It gives the lower bound of the decay time $\tau_{decay}^{min}$ of ~ 1000 nsec. The measured $I_{sub}$ values at three small staircases of Fig. 2b (~ 1.30 pA at $\Delta t = 100$ psec, ~ 2.62 pA at $\Delta t = 200$ psec, and ~ 4.76 pA at $\Delta t = 300$ psec) are converted to the current injected into a QD molecule per single pulse ($I_{pulse}^{QD}$, it is much larger than the detected current.) by the following formula;

$$I_{pulse}^{QD} = I_{sub}/(\Delta t/T_R)/(T_R/\tau_{decay}^{min})\ /(N_{dot}A_{dot}S_A), \qquad (1)$$

where $T_R$ is the pulse repetition time (= 10 nsec) and $S_A$ is the broadband sensitivity (~ $10^{-5}$) of the low noise preamplifier obtained from the frequency spectrum. The values of $I_{pulse}^{QD}$ at three staircases give the charge of approximately 70 % of 1, 2, and 3 electrons injected during $\Delta t$.

The scattering time by the acoustic phonon is calculated to be 25 nsec at T = 4 K for the QD energy separation of 1 meV[9]. This value is much longer than the single electron charging time as well as the coherence time, but it is comparable to the pulse repetition time of 10 nsec. A large value of $\tau_{decay}^{min}$ (~ 1000 nsec) suggests that only a small number of the electrons in the QD molecules are decayed before the next pulse after finishing coherent evolution at the end of the first pulse. In most cases, the other electrons are reset to the S state by the phonon scattering and the same evolution begins with the next pulse.

The observation of no clear oscillations at the second staircase (2 electrons injected) and the slight difference in the coherence time at the first (1 electron) and at the third staircase (3 electrons) might suggest that the time evolution of the quantum states and the resulted $I_{sub}$ should be a function of the number of injected electrons.

This work was supported by the Korean Ministry of Science and Technology through the Creative Research Initiatives Program.

**Correspondence and requests for materials should be addressed to D. A. (e-mail: dahn@uoscc.uos.ac.kr).**

**Figure 1** DC and pulsed transport of QD molecule. **a**. TEM bright field image of the QD molecule. Two stacked layers of InAs self-assembled QDs with the base width of 17 nm and the separation of 5 nm were grown by MBE. **b**. DC differential conductance dI/dV of the tunneling structure incorporating the QD molecules at high magnetic fields and at low temperature. It exhibits clear double spin splitting at high magnetic field. The spacing between two peaks before spin-splitting is converted to ~ 1 meV that is consistent with the symmetric/anti-symmetric (S/AS) energy separation $\Delta E$ of the coupled QDs. **c**. Schematic device structure for the pulse injection/probing measurements. The stacked InAs QDs were embedded in GaAs and Au pulse electrodes were deposited on top. The pulse electrodes were designed and tested for efficient transmission of short-period pulses to the wafer. The effective pulse height $V_{eff}$ at the electrode was calibrated by the sampling scope. The probe current $I_{sub}$ was measured from the $n^+$ substrate. **d**. DC transport of the pulse device. Both current and dI/dV show clear double peak. The value of $\Delta E$ estimated from the peak positions is also ~ 1 meV. **e**. Pulsed current-voltage characteristics. It shows similar characteristics as the DC transport.

**Figure 2** Pulse width $\Delta t$ dependence of the probe current $I_{sub}$. **a**. Schematic band diagram during the pulse injection. When the pulse is on and the pulse voltage is adjusted for the chemical potential of the Au being aligned in between the S/AS states of the QD molecule, an electron is injected into the QD molecule and the time evolution between two states begins. The system returns to the thermal equilibrium and the electron is decayed into the substrate and collected by the amplifier. The decay rates of the S and the AS state are different from each other and $I_{sub}$ changes depending on the final state probabilities. **b**. $I_{sub}$ vs $\Delta t$ and $dI_{sub}/d(\Delta t)$. They show plateau before strong increase when $\Delta t > 400$ psec. The large plateau consists of three small staircases whose value corresponds to the injection of single electron into the molecule.

**Figure 3** Observed oscillations of $I_{sub}$ ($\Delta I_{sub}$) as a function of $\Delta t$ at various regions of $\Delta t$. The Fourier transform of the data are plotted in the right column. **a**. $100 < \Delta t < 150$ psec. **b**. $290 < \Delta t < 340$ psec. **c**. $350 < \Delta t < 400$ psec. **d**. $290 < \Delta t < 340$ psec at the temperature T of 88 K. The oscillations are disappeared at high T. **e**. $350 < \Delta t < 440$ psec. There is a clear transition into the region of no oscillations. This transition at $\Delta t = 400$ psec occurs when $I_{sub}$ begin to increase rapidly. The periodicity $\Delta T$ observed in a, b, and c (~ 4 psec) gives the coherence energy ($h/\Delta T$) of ~ 1 meV which is the same as $\Delta E$ of the QD molecule.

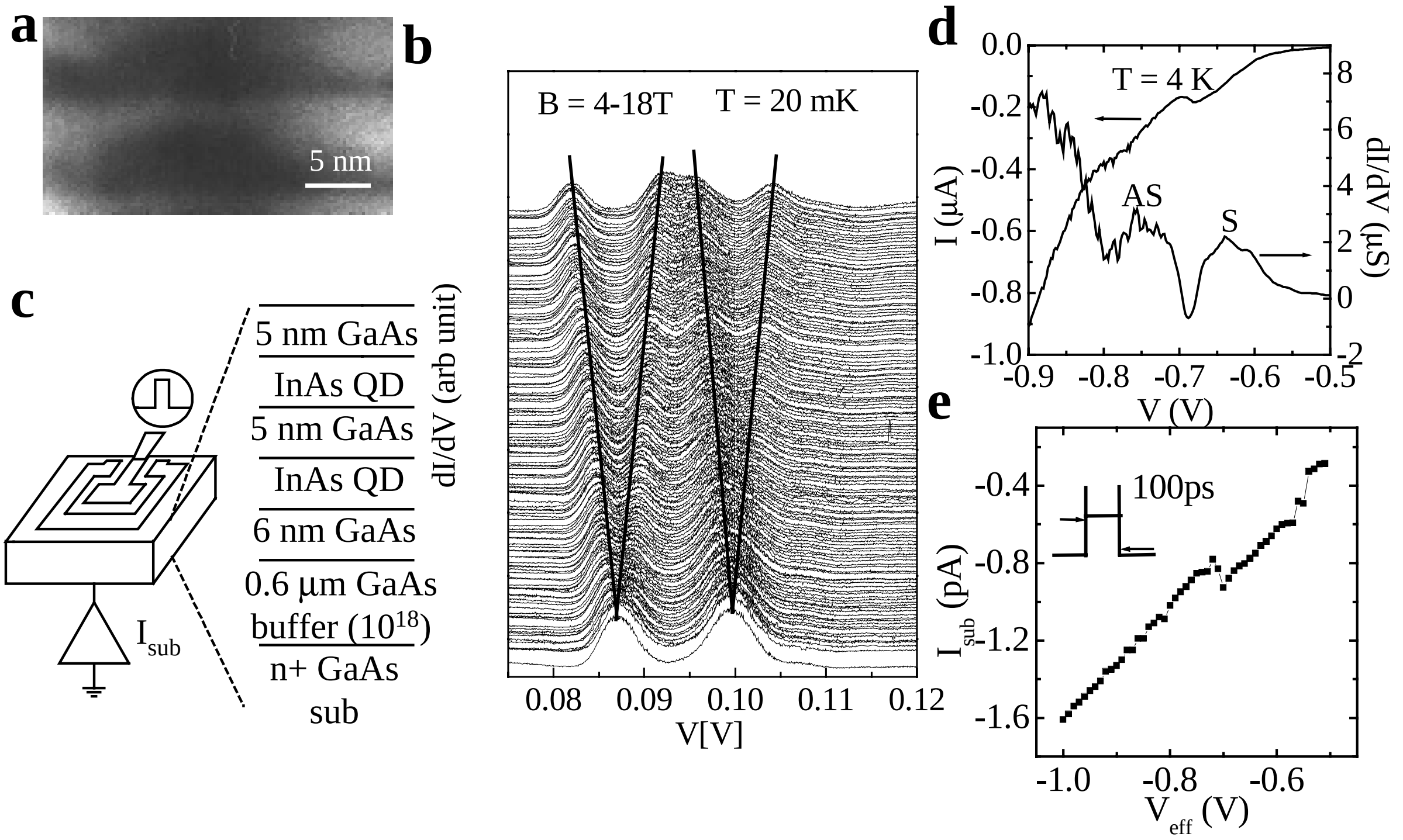
a
5 nm
b
B = 4-18T
T = 20 mK
dI/dV (arb unit)
0.08
0.09
0.10
0.11
0.12
V[V]
c
5 nm GaAs
InAs QD
5 nm GaAs
InAs QD
6 nm GaAs
0.6 μm GaAs
buffer (10^18)
n+ GaAs
sub
I_sub
d
T = 4 K
AS
S
I (μA)
dI/dV (μS)
V (V)
e
100ps
I_sub (pA)
V_eff (V)

Fig. 1

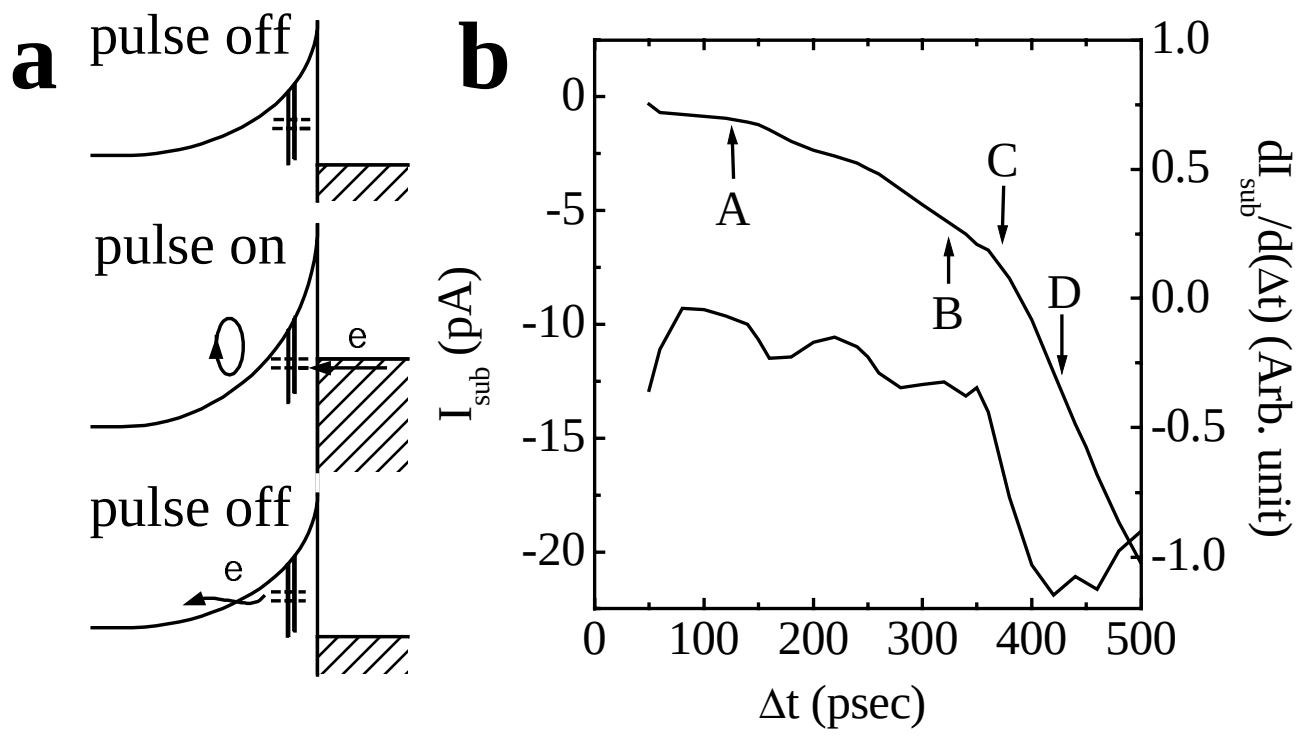
a
pulse off
pulse on
e
pulse off
e
b
A
B
C
D
0
-5
-10
-15
-20
$I_{sub}$ (pA)
1.0
0.5
0.0
-0.5
-1.0
$dI_{sub}/d(\Delta t)$ (Arb. unit)
0
100
200
300
400
500
Δt (psec)

Fig. 2

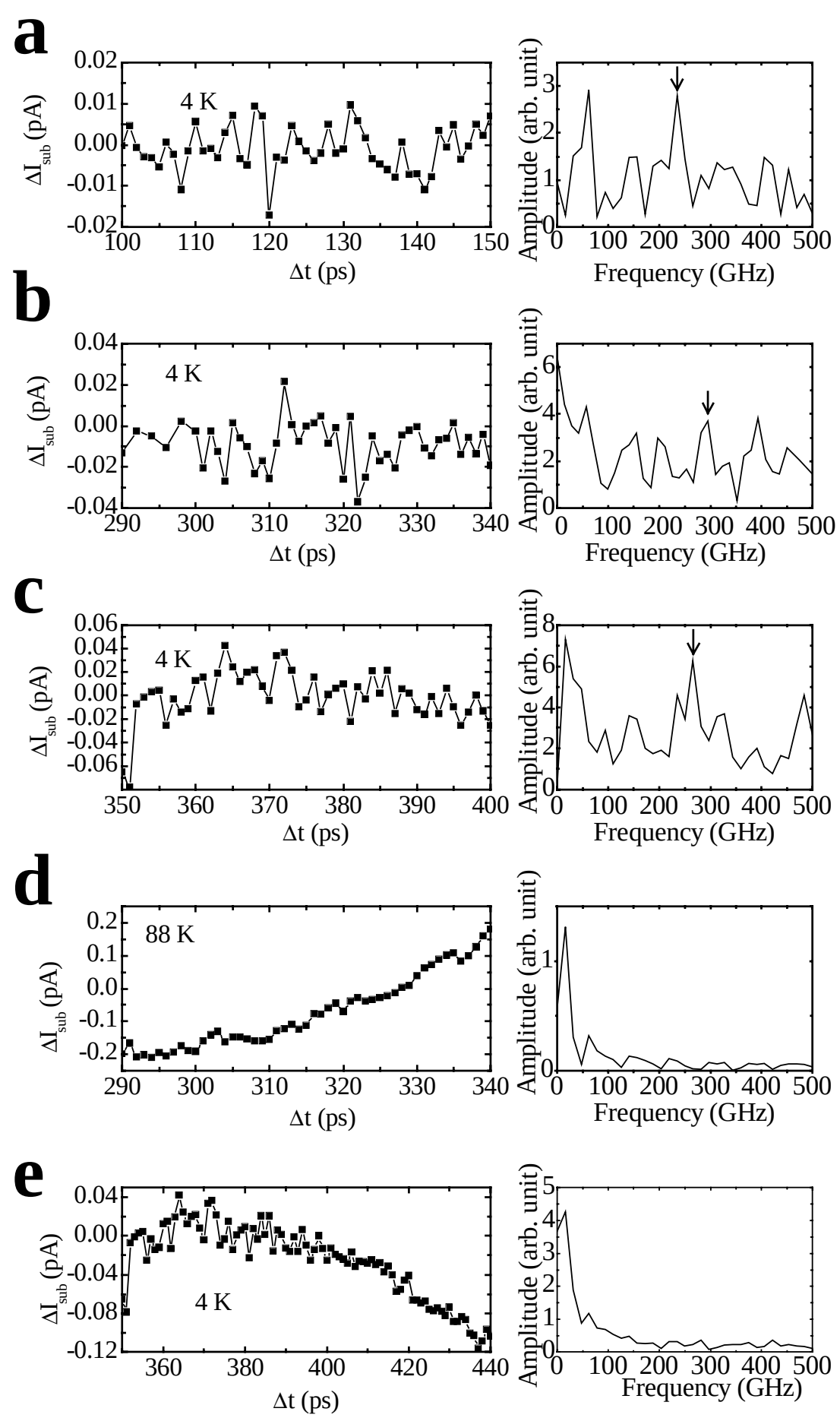
a
4 K
ΔI_sub (pA)
Δt (ps)
Amplitude (arb. unit)
Frequency (GHz)
b
4 K
ΔI_sub (pA)
Δt (ps)
Amplitude (arb. unit)
Frequency (GHz)
c
4 K
ΔI_sub (pA)
Δt (ps)
Amplitude (arb. unit)
Frequency (GHz)
d
88 K
ΔI_sub (pA)
Δt (ps)
Amplitude (arb. unit)
Frequency (GHz)
e
4 K
ΔI_sub (pA)
Δt (ps)
Amplitude (arb. unit)
Frequency (GHz)

Fig. 3